\documentclass[fleqn,10pt]{wlscirep}
\usepackage{braket}
\RequirePackage[colorlinks=true, allcolors=blue]{hyperref}

\newcommand{\be}{\begin{equation}}
\newcommand{\ee}{\end{equation}}
\newcommand{\br}{\begin{eqnarray*}}
\newcommand{\er}{\end{eqnarray*}}
\newcommand{\ba}{\begin{eqnarray}}
\newcommand{\ea}{\end{eqnarray}}
\newcommand{\bp}{\begin{minipage}}
\newcommand{\ep}{\end{minipage}}

\renewcommand{\u}{\uparrow}
\renewcommand{\d}{\downarrow}

\title{The interaction of excited atoms and few-cycle laser pulses}

\author[1]{J. E. Calvert}
\author[1]{Han Xu}
\author[1]{A. J. Palmer}
\author[2]{R. D. Glover}
\author[1]{D. E. Laban}
\author[3]{X. M. Tong}
\author[4]{V. K. Dolmatov}
\author[5]{A. S. Kheifets}
\author[1,6]{K. Bartschat}
\author[1]{I. V. Litvinyuk}
\author[1]{D. Kielpinksi}
\author[1,*]{R. T. Sang}
\affil[1]{Australian Attosecond Science Facility and Centre for Quantum Dynamics, Griffith University, Brisbane, Queensland 4111, Australia}
\affil[2]{Institute for Atomic and Nuclear Physics, University of Liege, Liege 4000, Belgium}
\affil[3]{Graduate School of Pure and Applied Sciences, and Center for Computational Science, University of Tsukuba, Tsukuba, 305-8571, Japan}
\affil[4]{Department of Physics and Earth Science, University of North Alabama, Florence, AL 35632, USA}
\affil[5]{Research School of Physics and Engineering, The Australian National University, Canberra ACT 0200, Australia}
\affil[6]{Department of Physics and Astronomy, Drake University, Des Moines, IA 50311, USA}

\affil[*]{corresponding author email address: R.Sang@griffith.edu.au}

\begin{abstract}
This work describes the first observations of the ionisation of neon in a metastable atomic state utilising a strong-field, few-cycle light pulse.
We compare the observations to theoretical predictions based on the Ammosov-Delone-Krainov (ADK) theory and a solution to the time-dependent
Schr\"{o}dinger equation (TDSE).  The TDSE provides better agreement with the experimental data than the ADK theory.
We optically pump the target atomic species and demonstrate that the ionisation rate depends on the spin state of the target atoms and provide physically transparent interpretation of such a spin dependence in the frameworks of the spin-polarised Hartree-Fock and random-phase approximations.
\end{abstract}
\begin{document}

\flushbottom
\maketitle
%
%
\thispagestyle{empty}

\section*{Introduction}

Recently, there has been much interest in the generation and utilisation of few-cycle light pulses that have a length of three or even less optical cycles.
This interest is in no small part due to the possibilities in applications such as lightwave electronics~\cite{Vrakking2006,Krausz2003}, high-order
harmonic generation~\cite{Brabec1998}, above-threshold ionisation, and multiple ionisation~\cite{Rudenko2004}.  Additionally, the precise control
of the carrier envelope phase (CEP) of a few-cycle laser pulse in strong laser--matter interactions opens many possibilities~\cite{deSilvestri2001,Tondello2003,Krausz2004}.
All these effects share a common starting point, namely, the strong-field ionisation of an atom.

Strong-field atomic ionisation is a highly nonlinear process that has been realised through high laser intensities obtained by tightly focusing
an few-cycle pulse of light with a high peak pulse power, even if the energy per pulse is relatively small~\cite{Krausz2000}.
The different interaction regimes that few-cycle light--matter interactions can be characterised by depend on the magnitude of
the electric field in the interaction region  relative to the ionisation potential of the atom.  In the first regime, the electric field is strong enough
to induce a perturbative non-linearity in the matter, but not strong enough to cause significant ionisation of atoms.  In the second regime, the electric field
is sufficiently strong to provide a high probability of ionisation in the target material.  This is known as the strong-field regime.  The Keldysh parameter, \(\gamma = \sqrt{{I_p}/{2U_p}}\),
is used to determine what regime a particular interaction belongs to.  Here \(I_p\) is the ionisation potential of the medium and
\(U_p\) is the ponderomotive energy, i.e., the kinetic energy imparted to an ionised electron by a linearly polarised oscillating electric field~\cite{Keldysh1965}.
The perturbative regime corresponds to \(\gamma > 1\) and the strong-field regime to \(\gamma < 1\)~\cite{Krausz2000}.

In the strong-field regime, it is possible to treat the light-matter interaction semi-classically through a three-step model~\cite{Corkum1993} that describes
the effects and results of the interaction.  The first step corresponds to ionisation by the light pulse as a result of the suppression of the Coulomb
atomic potential.  The second step involves the acceleration
of the electron wavepacket by the electric field of the light pulse.  The motion of the ionised electron corresponds to the classical motion of a charge
in an oscillating electrical field, which then imparts ponderomotive energy to the ionised electron.  The third step may include a recollision of the electron,
which results in a variety of interactions with the parent ion.

Inelastic recollision can result in secondary electron promotions within the parent ion, either causing a direct secondary ionisation known as non-sequential
double ionisation (NSDI)~\cite{Rudenko2004} or exciting another valence electron to a higher energy state.  This excitation lowers the effective second ionisation
potential of the atom, thereby providing the opportunity for ionisation in the remainder of the laser pulse in a process known as recollision-enhanced
secondary ionisation (RESI)~\cite{deJesus2004}.  The study of NDSI and RESI provides enlightening information about the electron dynamics of an ionising system.  Another possible interaction is the recombination of the wavepacket with the parent ion and the subsequent
creation of a photon that has a harmonic frequency of the driving field.  This process is known as high-order harmonic generation (HHG)~\cite{Krausz2000} and is being studied as a potential method to create a tabletop XUV laser source.
Elastic collisions may also occur, or the trajectory of the returning electron may not intersect with the parent ion.  In the latter case the atom remains singly ionised, but
a process known as above-threshold ionisation (ATI)~\cite{Corkum1989} may still occur.

As all these processes depend upon the initial ionisation, it is vital to have a good understanding of this process.
The most common theoretical method used to describe the process is based on the work of Ammosov, Delone, and Krainov.  It is commonly known as the ADK theory~\cite{ADK1986} and
makes two essential assumptions, namely: 1) Only the initial and final wavepackets of the electron are relevant in the ionisation process. 2) The energy of a single photon
is not sufficient to promote the valence electron into the continuum state, nor is the electric field of the peak high enough to suppress the atomic potential barrier
sufficiently to release the valence electron to the continuum.  The ADK theory yields an analytic expression for the tunnel ionisation rate.  In atomic units, which
are in use throughout this paper, it is given by
\begin{equation}
w\left(t\right) = C^2_{n*l *}\left(\frac{3|E\left(t\right)|}{\pi F_0}\right)I_p f\left(l,m\right) \left(\frac{2 F_0}{|E\left(t\right)|}\right)^{2n*-|m|-1}\textrm{exp}\left[\frac{-2F_0}{3|E\left(t\right)|}\right].
\label{ADKeqn}
\end{equation}
Here \(|E(t)|\) is the electric field of the laser pulse, \(n^*\) is the effective principle quantum number, \(l^*\) is the effective orbital angular quantum number,
\(m\) is the projection of the angular momentum quantum number, \(I_p\) is the ionisation potential of the target species, \(F_0 = \sqrt{2I_p}\), and \(C^2_{n*l*}\)
is a dimensionless constant that is unique for the atomic system under consideration.  Approximating a solution for \(C^2_{n*l*}\) was the purpose of the work done by
Ammosov {\it et al}.  Finally, the term \(f\left(l,m\right)\) is given by
\begin{equation}
f\left(l,m\right) = \frac{\left(2l+1\right)\left(l+|m|\right)!}{2^{|m|}\left(|m|\right)!\left(l-|m|\right)!}.
\label{lolf}
\end{equation}
The ADK theory is not valid in the intensity regime for over-the-barrier ionisation (OBI).  There have been several attempts to rectify this shortcoming.
One method involves correcting the wavefunction of the ejected electron for the Coulomb potential~\cite{Krainov1997}, thereby accounting for the possibility of OBI~\cite{Brabec1998}.
Another modification to account for OBI involves examining the ionisation rates across a broad range of atomic species, and then using the data to apply an empirical
correction to the ADK formula~\cite{Lin2005}.

Despite the limitations of ADK-based methods for calculating the ionisation rate, they are attractive to utilise since they are computationally far less expensive
than attempting to find solutions of the time-dependent Schr\"{o}dinger equation (TDSE) for the ionising system.  This has made ADK modelling the traditional method until
the past decade, when several techniques to obtain approximate solutions of the TDSE were developed (see, for example, \cite{tong2006,klaus2010} and references therein).
These techniques are taking advantage of significant increases in computational power and available resources.\

The aim of the present work is to investigate the strong-field ionisation from atoms in an excited state, for which there have been very few experimental investigations to date.
Experiments have been conducted to investigate the strong-field ionisation of Li~\cite{Schruike2011} (\(\gamma = 0.09\) to \(0.21\)).  That work, however, focusses on identifying the role of intermediate excited states
in the Li atom during the ionisation process, rather than considering ionisation from an initially excited atomic state as will be presented here.
Experiments examining the ionisation of metastable xenon have been performed by Huismans \textit{et al}~\cite{Huismans2011} using a \(\lambda = 7~\mu\)m laser
capable of providing pulses in the pico\-second regime in order to examine holography between directly ionised and rescattered electron wavefunctions (\(\gamma = 0.8\) to \(1.5\)).
Our work uses much higher laser intensities relative to the ionisation potential of the initial state than the work performed in~\cite{Huismans2011}
and investigates different ionisation regimes.  Recent experiments conducted by the authors demonstrate significant differences in the transverse
electron momentum distribution for the OBI regime compared to the tunnelling regime~\cite{Kheifets2015}. \

Singly excited states of noble-gas atoms have an electron in the valence shell, which leaves a hole in the remaining electron core.  The \(jj\) angular-momentum coupling scheme describes these states.  However, \(LS\) coupling notation suitably describes the $2p^5(^2P_{3/2})3s \ ^3P_2$ state of neon (hereafter defined as Ne\(^*\)) that we investigate in this work ~\cite{Khakoo1992}.
Ne\(^*\) is forbidden by selection rules to optically decay via single-photon dipole-allowed transition to the ground state.
It has a lifetime of approximately 14~seconds and has been previously used in laser cooling/atom trap experiments~\cite{Shimizu90,Vredenbregt2001,Matherson2007,Matherson2007a},
due to an accessible closed cooling transition to the $^3D_3$ state at 640.24~nm.  Below we present an experimental investigation of strong-field ionisation of Ne\(^*\).  Note that neon has a second metastable state (\(^3P_0\)) which is not considered in this work.\

Investigating the strong-field ionisation of a metastable noble-gas species is interesting for several reasons.  Firstly, the ionisation potential of Ne\(^*\) is only 5.1~eV, and hence
it is relatively straight forward to investigate OBI phenomena with laser systems that are routinely used in strong-field physics experiments.  Noble-gas species have closed single-photon dipole-allowed transitions that can be used to manipulate the trajectories of the
atoms as well as optically pump the target atom.  It is therefore possible to investigate the role of the initial atomic state in the strong-field ionisation process.
For example, it is possible to spin polarise the target atom and investigate ionisation dynamics from an orientated atomic system.\

Describing strong-field ionisation experiments is also a challenge to theory.  The critical field at which the unperturbed atomic energy level lie above the potential barrier and hence OBI
becomes possible is given by \(F_b = 4I_p^2/16Z_c\)~\cite{Lin2005}.  For Ne\(^*\), this corresponds to a laser intensity of \(2.7 \times 10^{12}\)~W/cm\(^2\), which is relatively low compared to the maximum available in our experiment.  Consequently, our experimental regimes can easily be varied from the case
where tunnelling ionisation is dominant to the case where OBI is the prevalent process.  This provides data from a challenging target over a wide range of experimental parameters
and facilitates an extensive test of our current theoretical understanding of strong-field physics.\

We present a new experimental apparatus that is capable of performing an experimental investigation of the strong-field ionisation of Ne\(^*\).
We compare the measured ionisation data to predictions from the ADK and the TDSE theories.  We also present first results for the ionisation of optically pumped Ne\(^*\) and
investigate the role of the initial atomic state in a strong field.

\section*{Results}

The experiment was prepared as described in the Methods section.  A number of data runs were performed at different laser intensities with Keldysh parameters
ranging from \(\gamma = 0.37\) to \(2.32\).  The experimental parameters were as follows.  The integration time of the experiment was 120~s.  The laser pulses had
random CEP and a pulse length of \(6.3 \pm 0.2\)~fs.  The measured atomic beam flux was \(1.4\pm 0.2 \times 10^{14}\)atoms/sr/s,
as measured by a Faraday cup detector.  In order to separate the Ne\(^*\)-only ion count from the \(^1S_0\) and \(^3P_0\) neon ion count, three separate measurements were taken.
The final Ne\(^*\) ion yield (\(S_{\rm Ne^*}\)) was determined according to
\begin{equation}
\label{signaleqn}
S_{\rm Ne^*} = S_{\rm coll-on} - S_{\rm coll-off} ,
\end{equation}
where \(S_{\rm coll-on}\) is the time-of-flight (TOF) measurement with the optical collimator on, and \(S_{\rm coll-off}\) is the TOF measurement with the optical collimator off. \(S_{\rm coll-off}\) contains ionisation information from all atomic states in the beam, while \(S_{\rm coll-on}\) contains information on an atomic beam with an enhanced Ne\(^*\) flux.  This results in ion yield information that is provided solely by the enhanced number of Ne\(^*\) atoms in the atomic beam.  Background contributions in all measured cases were less than 0.6\% of the signal.
Due to the low density of the atomic beam and a high vacuum of the COLTRIMS chamber,
it is assumed that all ions were created as a result of atom--laser-pulse interactions,
rather than any atom--atom interactions~\cite{Glover2012}.

The theoretical results were obtained as described in the Methods section.  In order to compare the predictions to experiment, the theoretical results were
scaled to fit the experimental intensity dependence using a Matlab two-parameter spline fitting procedure.  The scaling was done for both the ion yield and
the laser intensity using the equation \(y = A \times spline (\eta x)\).
Here \(spline\) is the spline function that is fit to the theoretical predictions, \(A\) is the ion yield scaling factor, and \(\eta\) is the laser intensity scaling factor.
The method has been used in previous work to compare theory to experiment in the case of atomic hydrogen~\cite{Pullen2013,Kielpinski2014}.  The uncertainty presented in the
experimental section is given by the Poissonian counting error.  Uncertainties in the laser intensity calibration include measurement error as well as systematic power
measurement to intensity calculation errors.  The latter is corrected with the intensity scaling.

\begin{figure}[!htb]
\centering
\includegraphics[width=\textwidth]{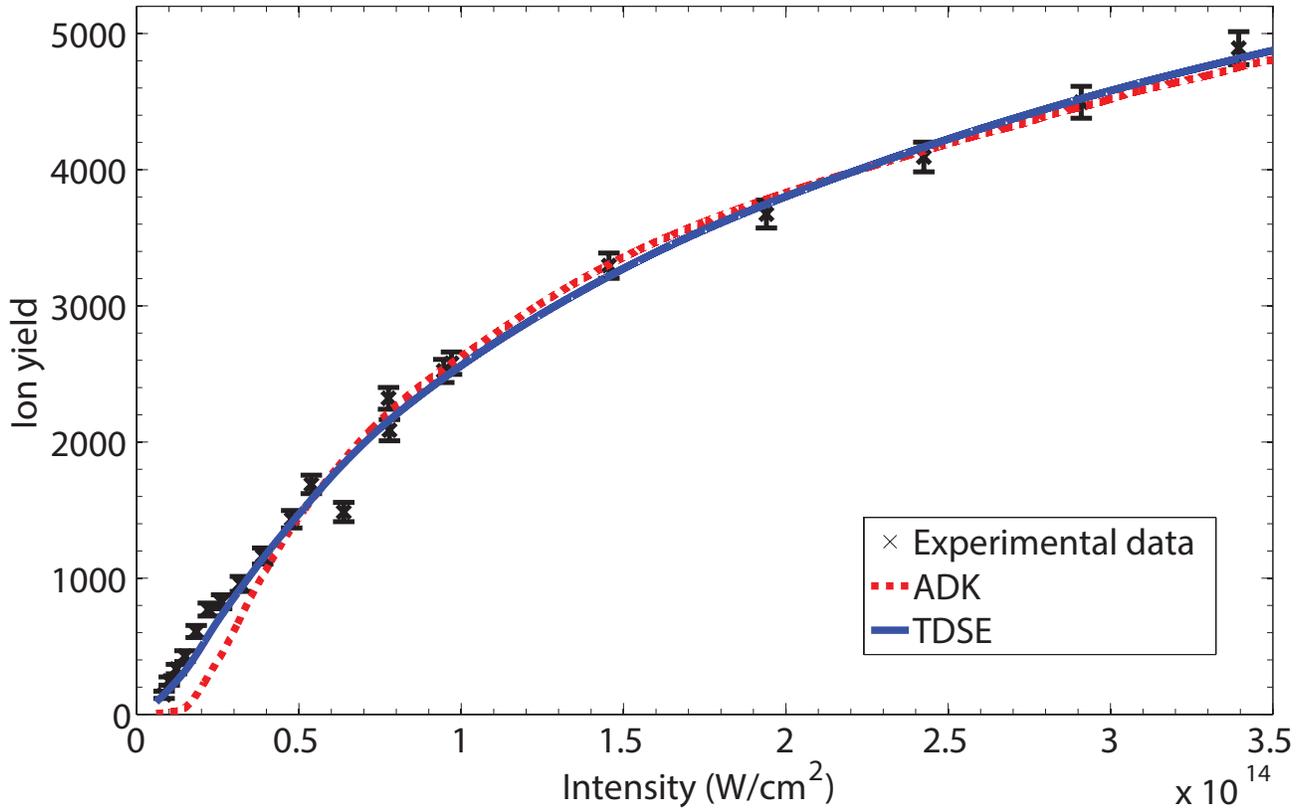}
\caption{(Color online) Comparison of experimental data with theoretical predictions.  The theories are scaled  using a spline fitting procedure.  For the ADK fit,
\(A = 0.18\) and \(\eta = 3.89\), with \(\chi^2 = 0.41\).  For the TDSE fit, \(A = 0.42\) and \(\eta = 1.59\), with \(\chi^2 = 0.25\).}
\label{yieldcompresults}
\end{figure}

It should be noted that there appears to be an outlying data point below the curve at \(6.38 \times 10^{13}\)~W/cm\(^2\).  The five data points at
\(6.38\), \(7.76\), \(7.79\), \(9.46\) and \(9.70 \times 10^{13}\)~W/cm\(^2\) were taken by employing two different experimental techniques for intensity
variation; one where the intensity was controlled solely by adjusting the half-wave plate and the germanium plates, and one where the intensity was
locked at the germanium plates while flip-in pellicle beamsplitters were used to reduce intensity.  This was done to determine the accuracy of
overlap between the two experimental techniques.  For the point in question it was ascertained that using the half-wave plate
for intensity control at this intensity would not effectively maintain the polarisation state of the light and it is likely an outlier caused by a
systematic error due to this issue.

\begin{figure}[!htbp]
\centering
\includegraphics[width=\textwidth]{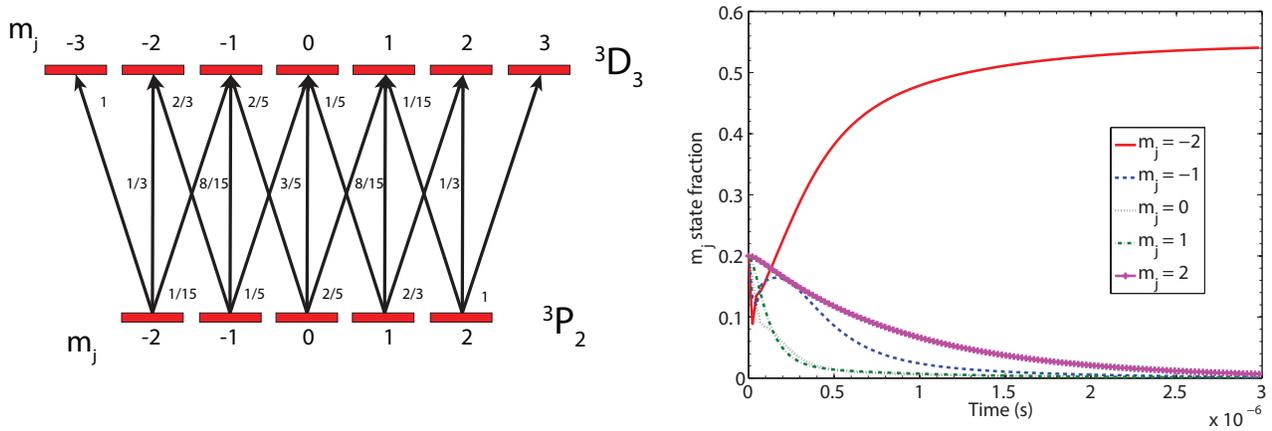}
\caption{(Color online) The panel on the left shows the optical pumping transitions for the \(^3P_2 \to 3^D_3\) states with the associated magnetic substates, also shown are the decay rates for the individual sublevel transitions.
The panel on the right displays the time evolution of the \(m_j\) states of \(^3P_2\) neon being pumped with \(\sigma^+\) polarised light tuned to
the \(^3D_3 \rightarrow ^3P_2\) transition.  The intensity of the light is 20 times the saturation intensity of the transition.
These results describe the system reaching steady state as described in the text, with 50\% of the atoms in the displayed \(^3P_2\) \(m_j = 2\) state.
The remainder exist in the \(^3D_3\) \(m_j = 3\) excited state, which is not displayed in the figure.  When the atoms leave the pump beam they
decay from the excited state into the \(m_j = 2\) state as described in the main text.}
\label{sigmapump}
\end{figure}

\begin{figure}[!htb]
\centering
\includegraphics[width=\linewidth]{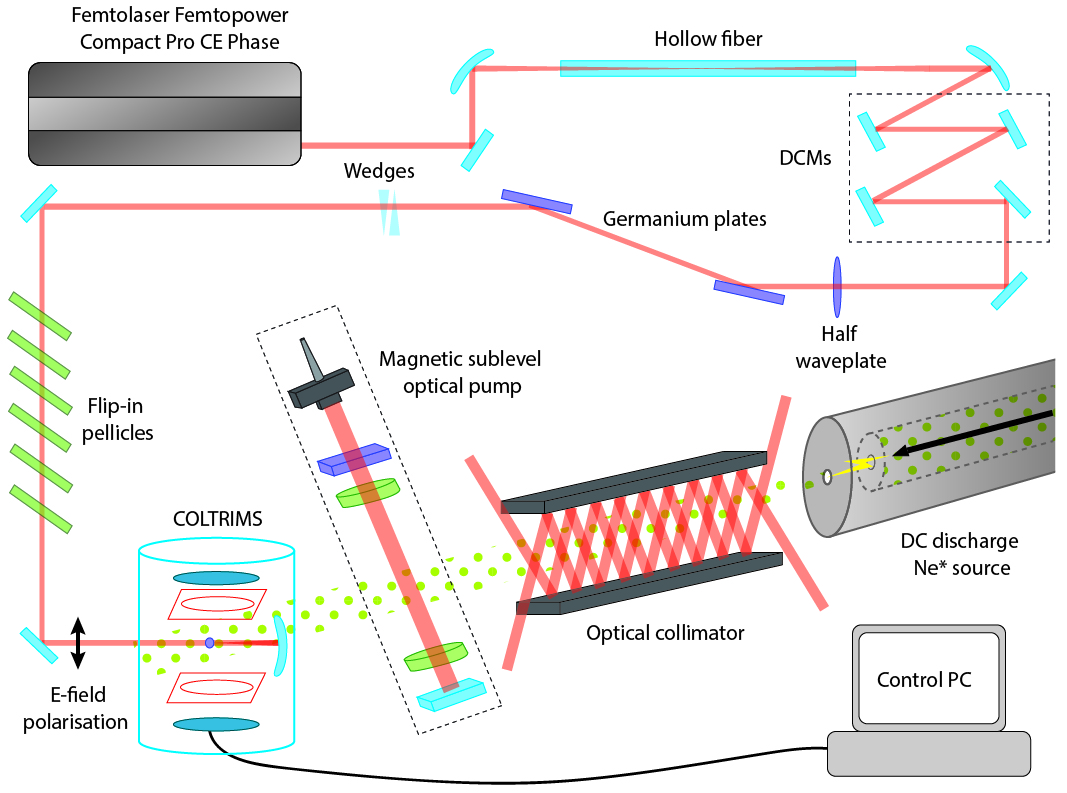}
\caption{(Color online) Schematic diagram of the experimental setup used in this work.  Only one pair of mirrors for the optical collimator is shown,
whereas two pairs are employed in the actual experiment to collimate in two directions.  The optical pump laser is propagating in the same direction as the electric field of the ionising laser, which defines the quantisation axis.}
\label{overallschematic}
\end{figure}

The present work also utilised optical pumping of the target atom with another laser beam tuned to the cooling transition in order to spin polarise the target atom.
If the optical pumping laser light is circularly polarised, it acts on a target atom by causing many single-photon absorptions followed by relaxation due to
spontaneous emission.  The result of this process is that the atomic population is transferred into the largest \(m_j = \pm 2\) states (\(+2\)
for \(\sigma^+\) left hand circularly polarised light and \(-2\) for \(\sigma^-\) right hand circularly polarised light) after the interaction with
the light beam \cite{varcoe1999}.
Atoms with these magnetic projection quantum numbers have the maximum total angular momentum and are spin polarised.  The sub\-level
transitions and their associated decay probabilities are shown in figure~\ref{sigmapump}.  An additional laser beam was added after the optical collimator to facilitate the optical pumping (see figure \ref{overallschematic}).  The laser beam interacted perpendicular to
the atomic beam and was on resonance with the cooling transition used in the optical collimator.  The laser beam was retro-reflected and the laser detuning
is set to 0~MHz so that the net scattering force on the atoms was zero~\cite{metcalf}, thus ensuring that the trajectory of the atoms remained unaltered, which avoided a loss in ion yield signal.  The polarisation state of this beam
was altered using a quarter-wave plate and facilitated a polarisation change which changed the distribution of \(m_j\) states.
The optical pump beam has a measured power of 125~mW, across a collimated beam geometry with a 6.1~mm radius.  This gives a pump intensity of 20~times
the saturation intensity (4.22mW/cm\(^2\)) of the optical transition.  We modelled the optical pumping process by numerically evaluating the optical Bloch equations (OBEs)
in the rotating-wave approximation (RWA).  The OBEs fully describe the evolution of the internal atomic states in the presence of an external field
including the atomic state coherences and spontaneous decay.  For example, figure \ref{sigmapump} shows the evolution of a Ne \(^3P_2\) atoms pumped
by \(\sigma^+\) light.  The system reaches a steady state after approximately 1\(\mu\)s with 50\% of the atoms in the ground \(^3P_2\) \(m_j = 2\)
state and 50\% of the atoms in the excited \(^3D_2\) \(m_j =3\) state.  A fully polarised state is only reached after a period of relaxation where
the system is allowed to evolve without the influence of the pump laser.  This second step takes a further 80~ns, after which approximately 99\% of
the atoms are in the desired \(^3P_2\) \(m_j = 2\) state.  On average, an atom was under the influence of the optical pumping beam for 12~\(\mu\)s,
which is more than sufficient to fully polarise the atomic beam.  Between the optical pumping region and the interaction region (approximately 45~cm),
there is a small residual magnetic field from the Earth, which could have induced a small depolarisation of the atoms.  However, our results show that
the majority of atoms remain well polarised.

\begin{figure}[!htbp]
\centering
\includegraphics[width=\textwidth]{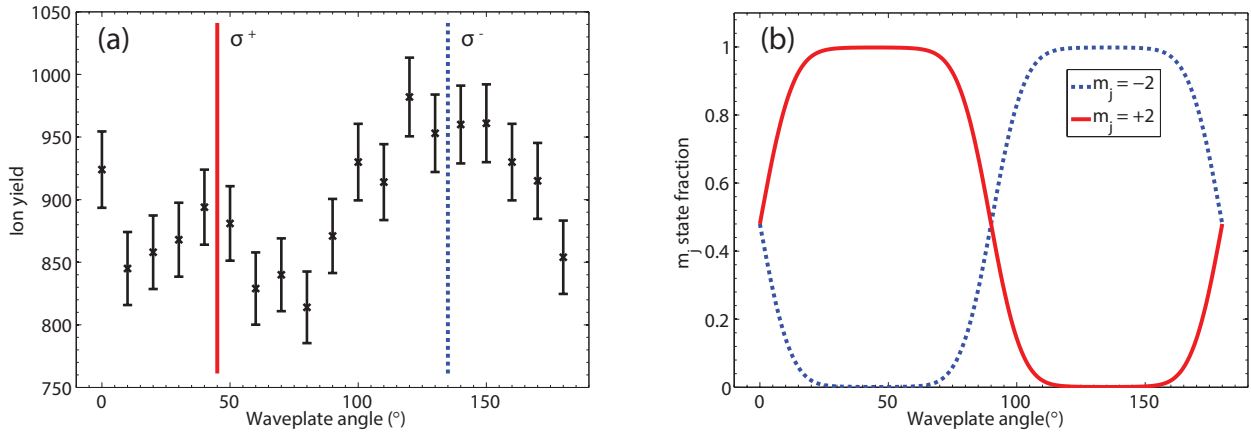}
\caption{(Color online) (a) Measurements of ionisation yield as a function of the angle that the quarter-wave plate makes relative to the polarisation axis defined by a linear polariser
when using 640.24~nm optical pumping light.  The intensity of the ionising laser is \(I = 9.2 \times 10^{13}\)~W/cm\(^2\).  The pump light is intended to pump
the atom beam into an ensemble of different \(m_j\) states, depending on the alignment of the fast axis of the quarter-wave plate with respect to
a linear polariser.  There is an \(8.7\%\) difference in ion yield between \(\sigma^+\) and \(\sigma^-\) pumped atoms, with an average error of \(4.8\%\) for
each data point.  (b) indicates the expected \(m_j\) state fraction of the beam at different wave-plate angles.  The modelling was performed
for the experimental pump beam parameters by numerically solving the OBEs and is provided as a guide for the eye.  For clarity, the remaining \(m_j\) states
are not displayed.}
\label{m_jpump}
\end{figure}

Figure~\ref{m_jpump} shows the ion yield when rotating the quarter-wave plate of the optical pump light.
As the pump light becomes more circularly polarised, there is a corresponding increase in the ionisation rate.  There are a number of important
observations that can be made about this measurement.  The change in the ellipticity of the optically pumping beam changes the atomic state
distribution of the Ne\(^*\) atoms, and we clearly observe an ionisation dependence on the initial state of the Ne\(^*\) system.
The second observation is an asymmetry in the ionisation distribution, where a higher ion yield is observed for the \(m_j = -2\) state.  This is a remarkable feature and was completely unexpected.

\section*{Discussion}

The results exhibited in figure \ref{yieldcompresults} are fit with arbitrary scaling and show overall good agreement with experiment for both the ADK and the TDSE theories.
Nevertheless, the predictions from the TDSE model show a qualitatively better agreement than the ADK at low intensities.  Ionisation due to OBI cannot be
predicted with ADK theory, so there is a high probability that the observed systematic underestimation of the ADK ion yield at intensities below
\(4.0 \times 10^{13}\)W/cm\(^2\) is a result of OBI ionisation experimentally occurring in that intensity regime.  As the TDSE solution accounts
for both OBI and tunnelling ionisation effects, one might expect to see a better scaled fit at those lower intensities.  This is indeed what we observe.
At higher intensities the fit after scaling is similar for both ADK and TDSE theories.  This is expected, as at near-unity ionisation
probability in the centre of the focus (i.e., saturation), focal volume averaging effects at the edges of the laser beam volume become the reason for
ionisation yield increase.  Since the ionisation yield depends upon the electric field amplitude, increasing the intensity of the laser above
the saturation point should yield a \(\sqrt{I_{pk}}\) dependence in the ion yield once the saturation point is achieved.  This is consistent with our dataset.

To obtain a semi-quantitative understanding of the experimental data in figure \ref{m_jpump}, we employ the concept of the spin-polarized Hartree-Fock (SPHF)
states~\cite{Slater} along with the spin-polarized random-phase approximation with exchange (SPRPAE)~\cite{JETP1983,JPB1993}.
The metastable state of Ne $2p^5(^2P_{3/2})3s \ ^3P_2$ is approximated by a superposition of two spin-polarized states with defined total spin projections:
Ne$^\dagger \ 2p^3_\u \ 2p^2_\d \ 3s_\d , m_s=0$
and
Ne$^{\star} \ 2p^3_\u \ 2p^2_\d \ 3s_\u , m_s=1.$
The so-defined spin-polarized ``daggers" and ``star" states serve as a convenient representation of the ionization process at hand.
Similar spin-polarized states have been used successfully to evaluate photoionization cross-sections of half-filled shells of transition
metal atoms~\cite{JETP1983,JPB1993,JPB88,JPB96,PRA2015}.  In the present case the gradual population of the \(m_s = 1\) state at the expense of the \(m_s = 0\) state describes optical pumping to the maximal \(m_j =2\) state of Ne\(^*\).  Indeed, $m_j=m_l+m_s=2$
necessitates both $m_l=1$ and $m_s=1$.

\begin{figure}[!htbp]
\centering
\includegraphics[width=8cm]{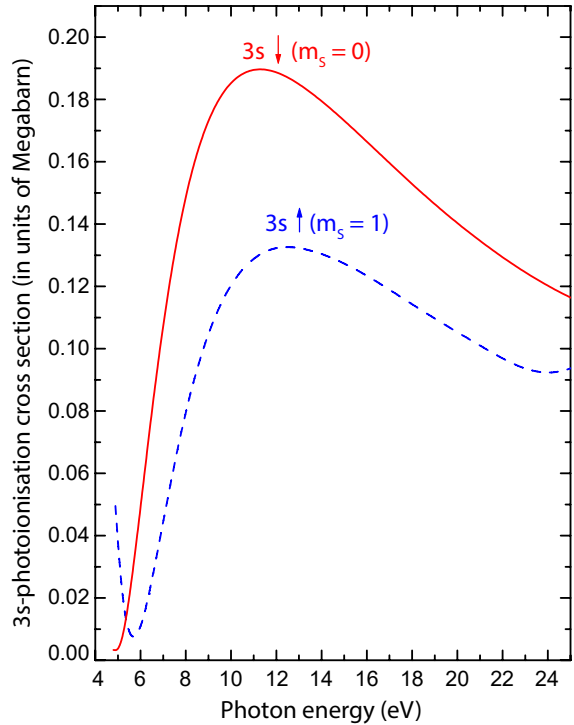}
\caption{(Color online) Calculated SPRPAE $3s_\d$ (red solid line) and $3s_\u$ (blue dashed line) photoionization cross sections of Ne$^{*}$.
The depicted cross sections are the geometrical averages of the results obtained in the
length and velocity gauges of the electric dipole operator, i.e., $\sigma = \sqrt{\sigma^{L}\sigma^{V}}$.
The calculated SPHF ionization potentials are $I_{3s\u}= 4.75$~eV and $I_{3s\d}= 4.66$~eV.}
\label{figNe3s}
\end{figure}

The binding energy and wave function of the $3s$ excited electrons in the two spin-polarized states are different.  Indeed, for the $3s_\d$
spin-down electron in Ne$^\dagger$, there \textit{is no} exchange interaction between this electron and \textit{three} $2p^{3}_\u$ core electrons.
In contrast, for the $3s_\u$ electron in Ne$^\star$, there \textit{is} exchange interaction between this electron and \textit{three} $2p^{3}_\u$ core electrons.
It is likely from the above discussion that the calculated photoionization cross section $\sigma_{3s\d}$ of the spin-polarized Ne$^{*}$ will differ
from $\sigma_{3s\u}$ of Ne$^\star$ already in the SPHF approximation. Inclusion of electron correlation into the corresponding calculations is known
to enhance the difference between photoionization of electrons with opposite spin polarizations compared to calculated SPHF results~\cite{JPB88,JPB96}.
In the present work, correlation in the $3s$-photoionization of Ne$^{\dagger}$ and Ne$^\star$ is accounted for in the framework of the ``spin-polarized''
random-phase approximation with exchange (SPRPAE)~\cite{JETP1983,JPB1993}.
The latter utilizes SPHF as the zero-order independent-particle basis -- the vacuum state and the corresponding infinite sequence of SPRPAE
inter-electron interactions are added to the photoionization amplitude~\cite{AC97}.

Figure \ref{figNe3s} depicts the calculated SPRPAE $3s$ photoionization cross sections $\sigma_{3s\d}$ of Ne$^{\dagger}$ and $\sigma_{3s\u}$ of Ne$^{\star}$.  They are found to differ from each other considerably. This is consistent with the experimental observation that
the optical pumping to the spin-up Ne$^\star$ state reduces the photo\-ionization cross section.
Note that the accuracy of the present calculation may be affected by the fact that the $2p^{2}_\d$ subshell is neither a half-filled nor
fully-filled subshell, whereas SPRPAE was originally developed for applications to atoms that have one or more half-filled sub\-shells
while the other sub\-shells are completely filled.  The present calculation is, therefore, term-averaged, and hence its accuracy may suffer.  Furthermore, the
calculation refers to the single-photon ionization process, whereas the experiment is performed in the strong-field tunneling ionization regime.

A comparison of the theoretical ADK ion yield for the \(m_j = 2\) and \(m_j =-2\) states using the two associated values for \(I_p\) given in the
caption of figure \ref{figNe3s} was performed.  The modelled laser intensity was taken the same as in figure \ref{m_jpump}.  ADK theory was employed as
it provides a simple, quick method of calculation when changing the ionisation potential.  The results indicated that the yield difference between an ensemble
of \(m_j =2\) and \(m_j = -2\) pumped atoms should be 9\% at I = \(9.2 \times 10^{13}\)~W/cm\(^2\).  These predictions were within the error of the experimental
results shown in figure~\ref{m_jpump}.

We have performed the first strong-field ionisation experiment with excited-state neon atoms and measured the complete ion yield from the ionisation of Ne\(^*\)
atoms using a COLTRIMS setup.  Our work showed that solving the TDSE, even with the necessary approximations to make the problem computationally
tractable, provides better agreement with experiment than the ADK theory.  This is likely the
result of applying the theories to an atom with such a low ionization potential, where the basic assumptions for ADK become invalid.
An 8.7\% difference in the ion yield was
experimentally demonstrated between atoms pumped to the \(m_j = -2\) state compared to atoms pumped to the \(m_j =+2\) state.  This is within
the error of ion yield predicted by a combination of ADK theory, SPHF, and SPRPAE.

\section*{Methods}
\subsection*{Experimental setup}
\label{exp_meth}
Few-cycle light pulses were provided by a commercially available chirped pulse amplification system (Femtopower Compact Pro CE Phase).
Following seed-pulse generation, pulse stretching, amplification, and pulse compression stages, the final laser output in typical operating
conditions was a 1~kHz train of pulses 6~fs long, with a pulse energy of approximately 450\(~\mu\)J.

The pulses from the laser system passed through a half-wave plate and a pair of germanium plates at Brewster's angle in order to
provide variable intensity from 150~mW down to 8~mW.  In order to preserve the polarisation state, a series of flip-in pellicle beamsplitters
were used to reduce the laser intensity further.  For intensity calibration purposes, a removable quarter-wave plate was also placed in the beam path.
The few-cycle pulses were then focussed into the interaction region of the detection system.  The intensity of the focussed light beam in the interaction
region was determined for a number of measured input powers, utilising the approach outlined in the work of Alnaser et. al.~\cite{Cocke2004}.  This method provided an absolute intensity
accurate to within~\(50\%\).  These data are used to create a calibration curve that maps the measured power to effective intensity.  In addition, this calibration
curve allowed for the calculation of the beam waist at the focus, assuming a Gaussian beam propagation.  The calculated beam waist diameter is \(14 \pm 1~\mu\)m,
with an associated Rayleigh range of \(810 \pm 120~\mu\)m.  The random shot-to-shot uncertainty of the laser intensity is dependent upon the uncertainty of the Thorlabs S310C power meter used to measure pulse power.  This was estimated to be \(11\%\) based on manufacturer specifications.

The detection system was a cold target recoil ion momentum spectroscopy (COLTRIMS) device.  This is an ultrahigh vacuum (UHV) system that utilises electric fields to
separate the products of a light-atom interaction based on charge polarity~\cite{Xu2013}.  The charged products are first guided onto multichannel plates to amplify the
signal, and then onto delay-line detectors that record the time and location of the ion strikes on the detector.  Electron momentum spectroscopy was available,
but not required for this experiment.  Mass spectroscopy of the product ions was performed by correlating TOF data from the delay line detectors.
This was used to obtain ion counts from the interaction with the ionising laser.  Position data can be used to determine ion momentum but was not utilised for this experiment.

A DC discharge source was used to generate the Ne\(^*\) atoms.  This type of source is common for generating metastable noble-gas atoms and was repurposed from previous
experiments~\cite{Sang2004,Baker2003}.  Neon gas was fed at a pressure of 1.1~Torr past a cathode tip and through a liquid nitrogen cooled 250\(~\mu\)m diameter
nozzle into the evacuated \((\approx 10^{-6}\)~Torr) source chamber.  The gas expands supersonically towards an anode skimmer, which provides collimation to
the atomic beam downstream.  The application of a high voltage across the two electrodes created a DC discharge in the region where the neon is expanding
into the vacuum system.  Electron collisions with neon atoms generate several products, including Ne\(^*\) at approximately 0.01\% efficiency~\cite{Gay1996}.

Immediately following the skimmer was an optical collimator~\cite{Hoogerland1996,Beardmore2009} which was utilised in order to increase the Ne\(^*\) flux.  The collimator
consists of two pairs of elongated mirrors at right angles to each other that are placed near parallel to the direction of travel of the atomic beam.
These mirrors were tilted slightly from parallel such that four incident laser beams frequency-locked close to the cooling transition for the \(^3P_2\) state create an angle frequency detuned two dimensional (2D) optical molasses along the path of the atomic beam.  This 2D optical molasses reduces the transverse velocity component of only the \(^3P_2\)
neon atoms and can be viewed as a collimation matter lens for the atomic beam.

Following the collimator, two chambers separated by two 1.5~mm apertures were used to create a differential pumping section in order to match the vacuum pressure to
the COLTRIMS UHV.  The first chamber contained electron deflector plates to remove charged particles from the atomic beam created by the discharge.
The second chamber contained a Faraday cup that is used to measure the beam flux and assisted in aligning the Ne\(^*\) source. A pneumatic gate valve separated
the Ne\(^*\) beamline from the COLTRIMS chamber.  Also on this chamber were a pair of optical viewports which allowed for the atomic beam to be illuminated perpendicular
to the atomic beam by two retro-reflecting laser beams at 640.24~nm, which are produced by a dye laser frequency locked and on resonance with the cooling transition.
A linear polariser and two quarter-wave plates were used to alter the ellipticity of the pump beam in order to pump the atoms into various \(m_j\) states.
When the atomic beam reached the interaction region of the COLTRIMS device, it had a diameter of \(1.5 \pm 0.3\)~mm, as measured by scanning the strong-field
laser beam focus across the atomic beam and observing the change in ion yield.

\subsection*{Modelling the ion yield}

\begin{figure}[!htbp]
\centering
\includegraphics[width=\linewidth]{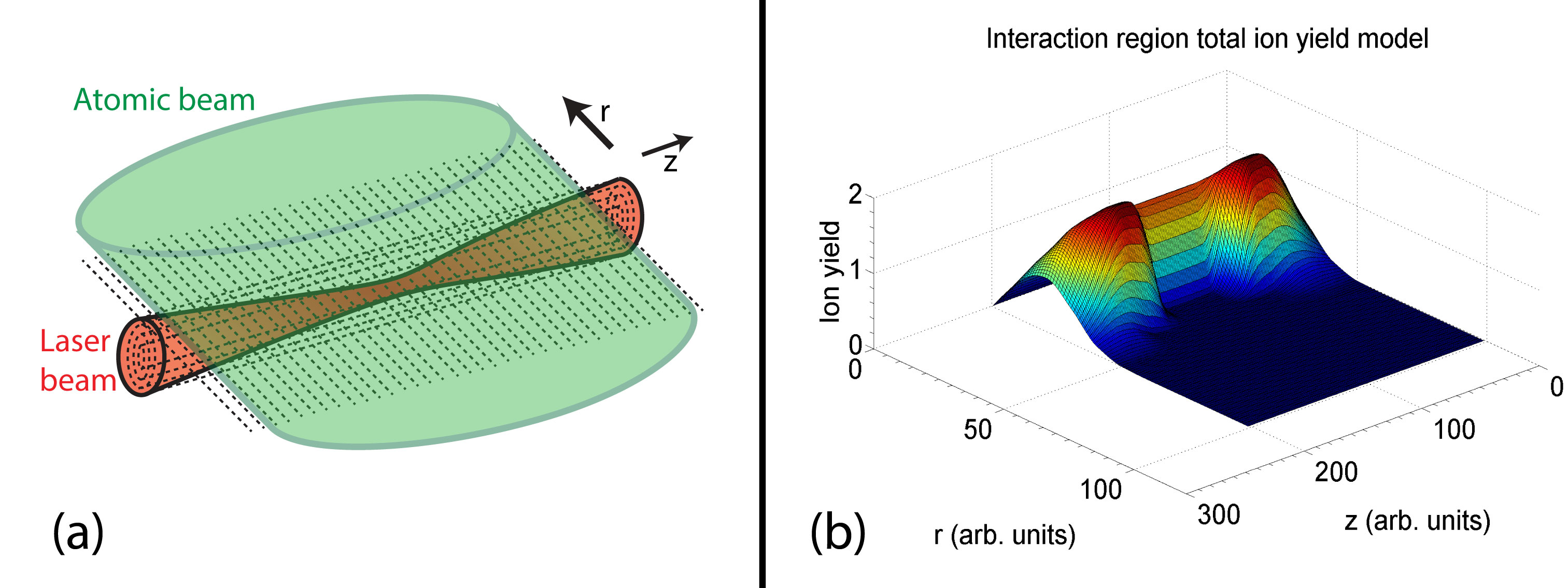}
\caption{(Color online) Part (a) is a schematic visualisation of the interaction region of the COLTRIMS. The atomic beam is travelling in the in the plane made with the z-axis and the \(\theta = 0\) angular coordinate.  As the system is solved symmetrically in \(\theta\), the axis along the \(\theta = 0\) coordinate is labelled the \(r\) axis as the solution requires knowledge of the displacement along the radial coordinate.  The laser beam is propagating in the \(z\) direction.  Part~(b) is a modelled 2D ionisation yield map for Ne\(^*\) interacting with a
laser pulse with the following parameters: \(I_{pk} = 9.6 \times 10^{13}\)~W/cm\(^2\); \(w_0 = 7.25~\mu\)m; \(T_{pul} = 6.3\)~fs; atomic beam
width = 1.5~mm; average atomic beam speed = 1000~m/s; atomic beam flux = \(1.4 \times 10^{14}\)atoms/sr/s.
These parameters, with the exception of \(I_{pk}\), were held constant throughout the modelling.}
\label{intregion_2dmap}
\end{figure}

In order to provide comparison to theory, a 3D focal-volume-averaged model was created and implemented through Matlab.  It is important to correctly
model the interaction region, since the low ionisation potential of metastable neon causes the ionisation probability to quickly reach unity at the
centre of the pulse at relatively low intensities.  This implies that, as the pulse intensity increases, the outer areas in the interaction region
significantly contribute to the total ion yield when compared to the ionisation of ground-state neon.  The model made the assumptions that the
laser pulse was Gaussian, the divergence of the atomic beam was negligible over the interaction region, the laser pulse was completely linearly
polarised, and all ions generated by the interaction were detected by the COLTRIMS.  Smoothing functions based on the work of~\cite{Kielpinski2014} were employed.
A representation of the interaction region is displayed in figure \ref{intregion_2dmap}, with axes labelled according to a cylindrical coordinate system.

In order to perform the focal-volume averaging, the cylindrical symmetry of the region was exploited.  This allows the interaction region to be flattened
into a 2D area that mapped the ionisation probability as a function of position in the interaction region. The atomic density was flattened
into a 2D area in the same way.  At this point, the density data was combined with the ionisation probability map to determine the total estimated
atomic ion yield at every point in the interaction region. These data were then integrated to give a total count for atoms ionised by a single pulse in
the interaction region.  Figure~\ref{intregion_2dmap} shows a typical 2D ionisation map generated by the script software.  At this point, the total
ion count result was multiplied by the number of pulses that are being modelled to give a final ion yield result for a laser pulse of any given intensity.

In order to generate a curve of ion yield as a function of intensity, a batch script was designed that creates a number of input peak laser
intensity \(I_{pk}\) values.  The script ran the ion yield script for each value of \(I_{pk}\) and generated a plot when the batch script was completed.
Two theoretical ion yields as a function of intensity plots were created for Ne\(^+\) ions.  One curve is generated by utilising ADK theory to
provide the ionisation probability as provided in Eq.~(\ref{ADKeqn}).  Values for \(C_{n^*l^*}\) were calculated by determining the wavefunction, \(\Psi^m\),
and the orbital energy of the Ne $3s$ atom~\cite{tong1997b}, before fitting to the expression~\cite{tong2002}
\begin{equation}
\label{ADKfitC}
\Psi^m\left(\mathbf{r}\right) = \sum_l C_l F_l(r)Y_{l m}( \mathbf{\hat{r}}).
\end{equation}
Here \(F_l(r)\) is the wavefunction in the asymptotic region where tunneling occurs, and \(Y_{l m}\) are spherical harmonics.\

Theoretical predictions for ionisation based on solving the TDSE are processed in the same manner.  The ionisation probabilities of the Ne \(3s\)
orbital were calculated by solving the TDSE under the single-active electron approximation with the second-order split-operator method in
the energy representation~\cite{tong1997a,tong2006}.  The model potential~\cite{tong2005} was calculated by using density functional
theory with a self-interaction correction~\cite{tong1997b}.  The calculated atomic ionisation potentials were in good agreement
with the measured ones.  The numerical convergence was cross-checked by comparing the ionisation probabilities obtained from the
integration of the ATI spectra and the survival probability of the $3s$ orbital as well as the excitation to other bound states.
The two results agree within a few percent.\

\section*{Acknowledgements}
The work of V.K.D.\ and K.B.\ was supported by the United States National Science Foundation under grants No.~PHY-$1305085$ and PHY-1430245.
We acknowledge support of the Australian Research Council in the form of the Discovery Project DP120101805 and DP110101894,
and of the ARC Centre of Excellence for Coherent X-ray Science.  JEC was supported by an Australian Postgraduate Research Award.

\section*{Author contributions statement}
J.E.C. assisted with devising the experiment, performed the experiment, experimental modelling and prepared the manuscript.
H.X., A.J.P., R.D.G., D.E.L. and I.V.L. assisted with the experiment.  X.M.T. provided ADK and TDSE theoretical data.
V.K.D. and A.S.K. provided physical interpretation of spin dependence on ion yield.  K.B., R.D.G., D.K. and R.T.S. provided theoretical support.
R.T.S. devised the experiment.  All authors contributed to writing of the manuscript.

\section*{Additional information}

The authors declare no competing financial interests.

\end{document}